# ALICE Detector Status and Commissioning

Hans-Åke Gustafsson, For the ALICE Collaboration

*CERN, Geneva, Switzerland and Lund University, Sweden*

**Abstract.** The Large Hadron Collider (LHC) will start operation in the end of 2007 colliding proton and lead beams at $\sqrt{s}$ = 14 TeV and $\sqrt{s_{NN}}$ = 5.5 TeV, respectively. The accelerator and the experiments are under construction and detailed studies of the physics program are being prepared. I will in this paper review the current status of the ALICE experiment and the heavy ion physics aspects that are unique at LHC.

## INTRODUCTION

Studies of matter under extreme condition have since long been an active area of research in physics. Starting from ordinary atomic systems to nuclei to systems composed of nuclear matter. Great interest in nuclear matter under extreme condition of high density and temperature was shown even before the theory of strong interaction, QCD, was established. Suggestions of the existence of new phases of nuclear matter were expressed and that these could be related to changes in the structure of the vacuum. The first studies of high temperature and density matter were performed about 25 years ago and the term "Quark-Gluon Plasma" (QGP) was suggested to describe this state. QCD, the theory of strong interaction, provides quantitative estimates of the critical temperature and density at which the phase transition from hadronic to quark matter should occur. Once established, the QGP provides a unique laboratory to study bulk properties of quark matter as well as the fundamental interaction of coloured objects in a coloured medium.

ALICE (A large Ion Collider Experiment) is the LHC experiment mainly focusing on heavy-ion physics. The ultimate goal in the field of relativistic heavy-ion collisions is to create and study the properties of the QGP. In addition to the heavy-ion program, ALICE will make use of the p-p running at LHC to collect reference data but also to pursue a p-p physics program complementary to the studies by the ATLAS and CMS experiments. ALICE will during the initial phase of LHC running collect p-p data. The 2008 LHC running will, besides p-p collisions, include a long heavy-ion run although not yet at full luminosity. Plans for the years after including p-A, light ions and different energies have been developed.

The ALICE collaboration has about 1000 collaborators from 80 institutes in 30 countries worldwide. 7 new institutes have joined the collaboration during last year and an application from US on joining the collaboration will be submitted in October.

The ALICE detector has been designed to measure at midrapidity ($|\eta| < 0.9$) most of the particles emitted in heavy-ion collisions. These measurements include identification and momentum determination with high precision. Hadrons with long lifetime will be identified by use of energy-loss and time-of-flight measurements while hadrons with short lifetime will be identified through their decay products. The momentum of the emitted charged particles is determined through tracking in a magnetic field ranging from 0.2-0.5 T. The full suit of detectors in the mid-rapidity region is used to achieve this. These detector systems are designed to cover a wide range in momentum, from very low values (100 MeV/c) to rather high values (100 GeV/c). This broad range makes ALICE unique in studying both soft and hard phenomena in heavy-ion as well as p-p collisions. The central tracking systems are complemented by a few systems for measurements of specific signals such as o'nium (J/$\Psi$, $\Upsilon$) states, photons, high momentum identified particles and global aspects of the collisions. The big challenge ALICE has to meet is to perform high precision measurements in an environment of extremely high particle densities which could go up to 8 000 particles per unit of rapidity. This corresponds to about 15 000 particles in the acceptance ($|\eta|<0.9$) of the ALICE central detectors. The ALICE detector performance has been optimized for 4 000 particles per unit of rapidity and checked up to 8 000 particles per unit of rapidity. A general discussion of the physics aspects of high-energy heavy-ion collisions can be found in [1]. In the following paragraphs the status of the different detector systems of the ALICE detector will be discussed.

## PROPERTIES IN THE LHC REGIME

PbPb collisions, with a center-of-mass energy about a factor 30 higher than at RHIC, will provide a qualitative new environment for studying nuclear matter under extreme conditions. The higher energy will improve by large factors all parameters relevant for the formation of the Quark Gluon Plasma (QGP). The created system will be hotter, bigger, denser and longer lived. The initial temperature will largely exceed the critical temperature predicted by lattice QCD [2] for QGP formation. The expected multiplicity in central PbPb collisions at LHC, based on extrapolation from RHIC data, will be between 1500 - 6000 charged particles per unit of rapidity. This extrapolation is affected by large uncertainties since the particle production at LHC will be dominated by hard processes, and the energy dependence for these processes in nuclear collisions is not very well known. The higher energy available at LHC opens up a new domain of small Bjorken x ($x = 10^{-2} - 10^{-5}$) characterized by high density saturated gluon distributions. Hard processes will dominate in the collisions. Low energy jets ($p_t < 10$ GeV/c) will be produced with very high rates and they will dominate the bulk properties of the colliding system. Measurements of these low energy jets may be important for understanding the thermal evolution of the system. Very hard probes (jets with $p_t >> 10$ GeV/c), abundantly produced, will make it possible to perform detailed studies of their interaction with the dense medium as well as studies of the jet fragmentation function. Weakly interacting probes like direct photons and gauge bosons are produced at rates sufficient to study jet tagging and nuclear parton distribution functions at very high $Q^2$.

In addition to the new information provided by the hard probes, the very high multiplicities expected at LHC, will allow for measurements on an event-by-event basis such as particle composition, spectra, flow and non-statistical fluctuation connected to critical phenomena.

## STATUS OF THE DETECTOR SUBSYSTEMS

The ALICE experiment is in the process of being assembled in the P2 cavern of the LHC inside the LEP L3 magnet. A schematic view of the detector is shown in figure 1.

### Inner Tracking System, ITS

The ITS has multiple purposes, one being an integrated part of the central tracking system to provide high resolution (< 100 μm) measurements of the primary vertex position as well as secondary vertexes for identification of short-lived particles. Another being a standalone system for measurements of low momentum particles (< 100 MeV/c). To cope with the extremely high particle density anticipated in heavy-ion collisions, a special ITS design is required.

The two innermost layers of the ITS are made of Si Pixel Detectors (SPD). They will provide good spatial and two-track resolution in the bending plane which is required in the high-multiplicity environment of heavy-ion collisions. The basic unit of the SPD is called a ladder which contains one Si sensor and bonded to 5 pixel chips. The chips are produced in radiation tolerant 0.25 μm CMOS technology. Five out of 10 SPD sectors are produced and under test.

The two next layers of the ITS system consisting of Si Drift Detectors (SDD) is mounted on a ladder structure holding 6 respectively 8 modules. The sensors contain the high-voltage divider which shapes the drift and collection fields. Six out of 36 SDD ladders are produced and tested and the mounting of the ladders on the support structure has started.

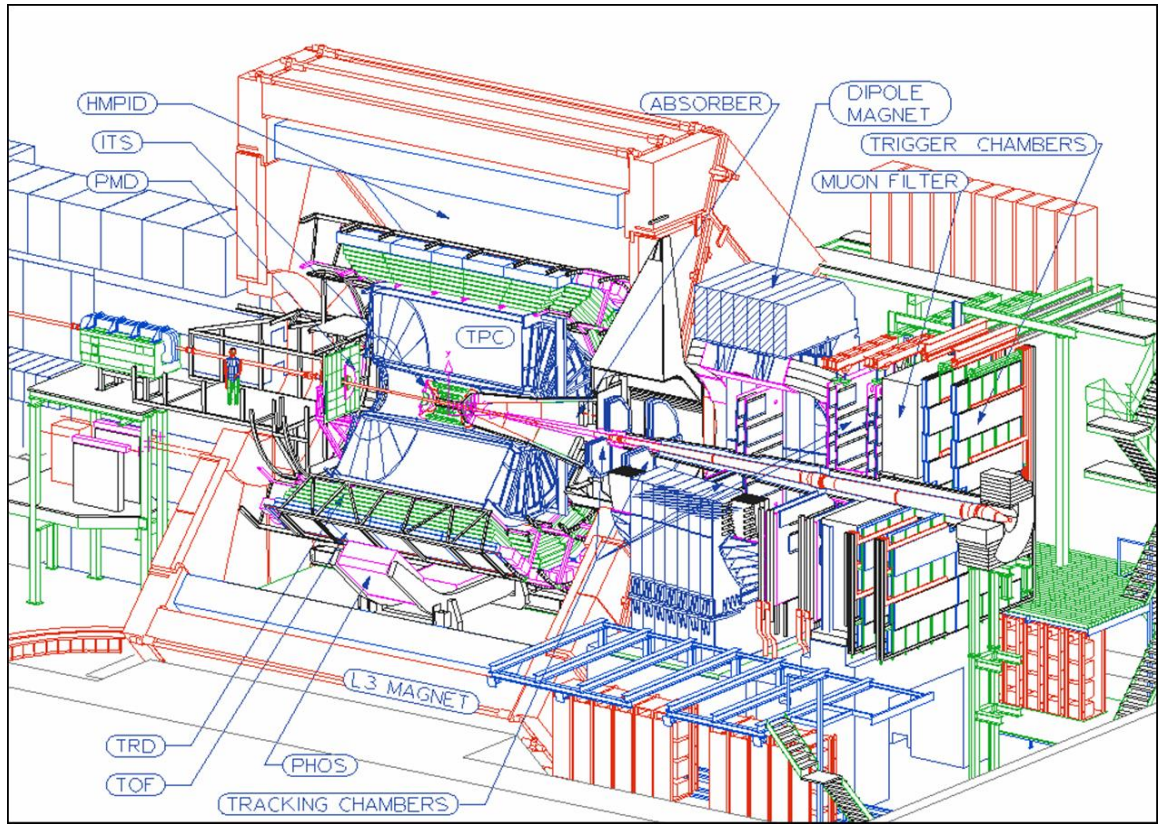

**FIGURE 1.** A schematic view of the ALICE detector.

The two outer layers of the ITS are composed of double sided Si Strip Detectors (SSD). They are mounted in a ladder structure similar to the SDD. The ladder frames are made of extremely thin and complex carbon-fiber structures to minimize the amount of material. These frames support the SDD ladders, the services for the SDD and SSD and the silicon rings of the FMD. All 72 SSD ladders are produced and the mounting on the support structure has started.

The production, assembly and testing of all three ITS systems is progressing well, but on a tight schedule, and they are expected to be ready for installation according to the global installation schedule.

## Time Projection Chamber, TPC

The TPC is the main tracking device of the ALICE detector. It is capable of measuring with high precision the momentum of all charged particles below 10 GeV/c and together with the other tracking detectors provide good resolution up to 100 GeV/c. Particle identification in the TPC requires good energy loss measurements which will depend on the track multiplicity inside the TPC. The resolution in the energy loss measurements is predicted to be about 5-6 % at a particle density below 2000 per unit of rapidity increasing to about 7 % at 4000 particles per unit of rapidity.

The TPC is constructed of 4 cylindrical vessels 5 m long. The active drift volume of the TPC is defined by the two field cage vessels which cover radii from 0.9 m to 2.5 m, respectively. The two containment vessels provide isolation of the TPC high voltage. The readout of the TPC is done using conventional multiwire proportional chambers with cathode pads. The signals from the cathode planes are processed in custom designed Front End Card (FEC). The production and installation of all components for the TPC is finished and commissioning with cosmics is in progress and the performance is according to the TDR specifications. The TPC will be transported down into the cavern in mid October to be installed in the L3 magnet.

# Particle Identification Systems

Particle identification over a broad range of momenta is very important for most of the observables anticipated in ALICE. The central tracking system of ALICE provides through energy loss measurements in the TPC and ITS discrimination of π/K/p in the non-relativistic regime. The vertex finding capability of the central tracking detectors gives identification of short-lived particles through their hadronic decays. Complementary to the central tracking system, ALICE has several other particle identifying detector systems like, TOF, TRD, HMPID and PHOS. The TOF systems provides separation for π/K/p and electrons with $p_t > 1$ GeV/c are identified and discriminated against charged pions in the TRD system. The Ring Imaging Cerenkov detector (HMPID) extends in a limited solid angle the PID capability of ALICE towards higher particle momenta, π/K up to 3 GeV/c and K/p up to 5 GeV/c. The PHOS detector system identifies photons in a limited solid angle by a combination of electromagnetic calorimetry and charged particle vetoing. A dedicated spectrometer in the forward direction identifies muons.

*Transition radiation detector, TRD*

Combining the information from the TPC, ITS and TRD will provide identification of high momentum electrons as well as discrimination against pions. The TRD will also operate as an electron spectrometer to measure charm and beauty through their semi-leptonic decays and o'nium states in their $e^+e^-$ decay channels. The fast tracking capability of the TRD will be used to trigger on high momentum electron- and hadron tracks which will be of interest for the jet-physics program. The first TRD super module is being assembled to be installed in September. Additional modules will be installed in the spring of 2007.

*Time-Of-Flight Detector, TOF*

Outside the TRD in the central barrel of ALICE is the TOF system which is built on Multi-gap Resistive-Plate Chambers (MRPC) technique. This system will together with the information from the ITS and TPC significantly improve the identification of pions, kaons and protons in the momentum range 200 MeV/c to 2.5 GeV/c. To achieve this, a time resolution of about 100 ps is required. Two TOF super modules are constructed and tested and they will be installed in September. Additional 9 super modules are under construction to be installed in the spring of 2007.

*High Momentum Particle Identification Detector, HMPID*

The HMPID based on ring-imaging Cerenkov technique has a liquid radiator and a multi-wire proportional chamber with pad readout. A thin layer of CsI is evaporated on the pad plane. The HMPID, with a limited coverage in rapidity ($-0.6 < \eta < 0.6$) and azimuth ($57.6^o$) is located 5 m from the interaction point. The purpose of this detector system is to extend the range of identifying and separating charged hadrons towards higher $p_t$, 3 GeV/c for π/K and 5 GeV/c for K/p. All 7 modules have been produced and installed in its support frame. The whole assembly will be installed in the cavern in September.

*Photon Spectrometer, PHOS*

The PHOS detector placed 5 m from the interaction point covers a limited part in rapidity ($-0.12 < \eta < 0.12$) and azimuth ($100^o$). PHOS is a highly segmented $PbWO_4$ calorimeter with the purpose of performing high resolution photon measurements and to discriminate against leptons and charged hadrons. The readout of PHOS is using low-noise APDs. A detector system based on multi-wire proportional chambers is placed in front of the PHOS spectrometer acting as a veto detector for charged particles. The high segmentation of the PHOS spectrometer makes it possible to identify neutral mesons through their two-photon decay channel. Using time-of-flight, shower-topology and isolation technique will provide additional particle identification especially to discriminate neutrons and anti-neutrons and identification of direct protons. Measurements of neutral pions and direct photons will be made over a broad range in momentum from about 1 GeV/c to about 80 GeV/c. The first full PHOS super module is presently being calibrated in beam and it will be installed in the experiment in October. Preparations for construction of additional 3 super modules have started.

## Muon Spectrometer

ALICE has in the forward direction a complex spectrometer for identification of muons. The main goal with this detector system is to perform high quality measurements of o'nium states with a resolution of 100 MeV/c$^2$ at the Y mass. The spectrometer consists of a complex hadron absorber, a 3 Tm dipole magnet, 5 planes of tracking chambers and 2 planes of trigger chambers. The dipole magnet has been successfully installed, commissioned and completely mapped. The installation of the trigger chambers is progressing well and the installation of the tracking chambers will start in September.

## Forward Detectors

ALICE will have various detector systems at large rapidities on both sides of the interaction point. These detector systems will provide information for triggering, event selection and global properties.

The Zero Degree Calorimeters (ZDC), located 110 m from the interaction point, are constructed of tantalum or brass with embedded quartz fibers. The neutron spectators will be measured by the tantalum systems while the proton spectators, deflected by the LHC dipole magnets, by the brass systems. The ZDCs will provide information for triggering and impact parameter determination. All calorimeter modules are produced and ready for installation in the LHC tunnel.

The Photon Mulitiplicity Detector, PMD, is located 3.6 m from the interaction point on the side opposite to the muon spectrometer. It is composed of proportional chambers sandwiching a passive Pb converter. The purpose of this detector system is to measure photons and charged particles at large rapidities to search for non-statistical event-by-event fluctuations and flow. All components of the system are in the production and the full system will be ready for installation according to the global installation schedule.

The Forward Multiplicity Detector, FMD, is composed of 5 discs of Si pad detectors, two of them are located on the muon spectrometer side and three on the opposite side. The FMD together with the ITS cover from -5.1 to 3.4 in pseudo-rapidity for charged particle measurements. The production, assembly and testing of all detector components is in progress and the systems will be ready to start installation in spring of 2007.

The T0 detector system, based on Cerenkov radiator + PMTs, consists of two arrays each having 12 elements. The arrays are placed one on each side of the interaction point, at 70 cm on the muon side and 350 cm on the opposite side. The purpose of the T0 detector system is to provide the start signal for the time-of-flight measurements as well as a signal for the L0 trigger with a time resolution better than 50 ps. The production of all the detector components is in progress and they will be ready for installation according to the global installation schedule.

The V0 detector system consists of two arrays of scintillators embedded with wavelength shifting fibres + PMTs. The arrays are placed on each side of the interaction point, at 90 cm on the muon side and at 355 cm on the opposite side. The purpose of the V0 systems is to provide the main interaction trigger but also to determine the vertex position online. It will also, through coincidence between the two arrays, be able to identify beam-gas interactions. The detector components are in production and they will be ready for installation according to the global installation schedule.

## Control Systems and Computing

### Detector and Experiment Control Systems, DCS and ECS

The DCS system will provide tools for an ALICE operator to have full control over all modes of operations of the experiment. It will allow the detector and service groups to have full control of their specific equipment in the experiment. It will also allow running the systems in a standalone mode during installation and commissioning. The DCS group has together with each detector systems written a user requirement document in which all requirements for the control of the systems have been defined. Most of the hardware and software for the DCS systems have been defined and several systems have run with full scale DCS system during commissioning. The DCS systems for all services and infrastructure will be installed and ready for the first LHC p-p collisions.

The ECS system is developed to coordinate all the activities in the DCS, DAQ, TRG and HLT systems. The architecture has been defined and the full system has been successfully used during the commissioning of several subsystems.

*Trigger, DAQ and HLT*

The ALICE trigger system delivers one pre-trigger and three trigger levels (L0, L1, L2). The T0 and V0 detector systems provide in less than 100 ns a pre-trigger with the purpose of sending a wake-up signal to the FEE of the TRD system. The L0 (1.2 μs) and L1 (6.5 μs) triggers are sent to the fast detectors while the L2 (100 μs) triggers the readout of the slow TPC detector. The Central Trigger Processor (CTP) and the Local Trigger Unit (LTU), used to distribute the triggers, are produced and have been used during commissioning of several detector systems.

The data are transferred in parallel to the Data Acquisition System (DAQ) over the Detector Data Link (DDL) via the Read Out Receiver Card (RORC) to a farm of individual computers called the Local Data Concentrators (LDC). The LDCs are building the sub-event that are sent to the final eventbuilder, the Global Data Collector (GDC) which has the capacity of processing in parallel 40 different events with a total bandwidth to storage of 1.2 Gb/s. The software framework of the ALICE DAQ, DATE (Data Acquisition and Test Environment), provides all the controls for the data flow. The DAQ system has been tested in Data Challenges with increasing complexity. The commissioning of both hardware and software is in progress and integration tests have been successfully performed during commissioning of several detector system.

The High Level Trigger (HLT) is a software trigger system which will be used for extracting rare event. Another very important task of the HLT system is to perform data compression. It will also perform fast partial or full reconstruction of the events which will be sent to the DAQ data flow at the LDC level. All these tasks will be done on a dedicated computer farm. The HLT system is fully defined and it has been operated during commissioning of several detector systems.

*Offline*

AliRoot is the ALICE offline-software framework based on the ROOT package. Tasks for reconstruction, event simulation and data analysis are all performed within AliRoot. This framework is constantly evolving to extend its functionalities. Interfaces to different external packages are developed to allow the users to switch between different packages. Interfaces for GEANT3, FLUKA and a selection of event-generators are in operation. Several physics data challenges have been performed using a substantial fraction of the final offline capacity. The main goals of the data challenges have been to produce and analyse about 10 % of the data taken in a standard data taking year, use the complete offline chain and to test the software and physics analysis of data. This is all entirely done on the GRID with the AliEn/LCG GRID services. The ALICE computing model was presented to the LHCC and the Computing Technical Design Report (TDR) [3] has been approved by the LHCC.

## Conclusion

All major detector systems in ALICE are in the production/installation phase. A detailed installation and commissioning plan has been developed to have the ALICE baseline detector ready for the first p-p collisions in the end of 2007. ALICE has developed a framework for producing the large number of simulated event needed for the preparation for the full physics analysis of data. The ALICE collaboration is looking forward to an exciting future of high energy heavy-ion physics in a new regime of physics as well as a dedicated p-p scientific program complementary to the ATLAS and CMS experiments.